 \documentclass{aastex}
 \usepackage{emulateapj5}
 \usepackage{apjfonts}
\shorttitle{CMB-NORMALIZED SZ PREDICTIONS}
\shortauthors{HOLDER}
\begin{document}
\submitted{Submitted to ApJL July 18,2002}
\title{%
 CMB-Normalized Predictions for Sunyaev-Zel'dovich effect fluctuations
}%

\author{Gilbert P. Holder}
\affil{%
 School of Natural Sciences, Institute for Advanced Study, 
 Princeton NJ, 08540; holder@ias.edu
}%

\begin{abstract}
 We predict the level of small-scale anisotropy in the cosmic 
 microwave background (CMB) due to the Sunyaev--Zel'dovich (SZ) 
 effect for the ensemble of cosmological models that are 
 consistent with current measurements of large-scale CMB anisotropy.  
 We argue that the recently reported detections of the small-scale 
 (arcminutes) CMB anisotropy are only marginally consistent with 
 being the SZ effect when cosmological models are 
 calibrated to the existing primary CMB 
 data on large scales. The discrepancy is at more than $2-2.5\sigma$, 
 and is mainly due to a lower $\sigma_8\la 0.8$ favored by 
 the primary CMB and a higher $\sigma_8\ga 1$ favored 
 by the SZ effect. A degeneracy between the optical depth to Thomson
 scattering and the CMB-derived value of $\sigma_8$ suggests that 
 the discrepancy is reduced if the universe was reionized very early,
 at redshift of $\sim 25$. 
\end{abstract}
\keywords{%
 cosmic microwave background --- cosmological parameters ---
 cosmology: observations
}%
\section{Introduction}\label{sec:intro}

The current generation of cosmic microwave background (CMB)
experiments are producing a wealth of information on a wide range of 
angular scales. 
With the recent measurements by BIMA \citep{dawson02} and CBI 
\citep{mason02} complementing earlier measurements 
\citep[see][for a recent compilation]{hu02}, CMB anisotropy has been 
measured over a range of multipoles of $l=2-6000$ 
(or angular scales of $2'-90^\circ$).

At low multipoles ($l \la 2000$) anisotropy is primarily
generated at $z\ga 1000$ except at very low
multipoles ($l \la 10$) where late-time decay of gravitational
potential contributes significantly.
At higher multipoles (smaller angular scales) low-redshift sources
generate a significant amount of fluctuation power. 
At the observing frequencies of CBI and BIMA ($\sim$ 30 GHz), 
the largest sources of low-redshift anisotropy are radio point 
sources and the thermal Sunyaev--Zel'dovich (SZ) effect.
The latter is of cosmological interest and may be large enough to 
be detected, depending on cosmology \citep[e.g.,][]{cole88}.

The reported detections of power at small angular scales
($l=2000-6000$) have argued that point-source contamination is not a 
problem, suggesting that the detected power could be due to 
the SZ effect \citep{bond02,dawson02,komatsu02}. 
Since the number density and brightness of the sources contributing to 
the SZ fluctuations (i.e., hot gas in galaxy clusters at $z \la 1$) 
depend on the background cosmology, the level of the SZ fluctuations depends on 
cosmological parameters.
The SZ angular power spectrum is sensitive to 
the matter-fluctuation amplitude and the baryon density 
of the universe but relatively insensitive to the matter density of the 
universe \citep{komatsu99} or other cosmological 
parameters \citep{komatsu02}. 
By fitting the CBI and BIMA data to theoretical predictions, 
\citet{komatsu02} have found a constraint on linear r.m.s. mass fluctuations
within an $8~h^{-1}~{\rm Mpc}$ sphere, $\sigma_8$, as 
$\sigma_8(\Omega_{\rm b}h/0.035)^{0.29}=1.04\pm 0.12$ at the 95\% confidence level.

On the other hand, observations of the primary CMB anisotropy and 
the large-scale structure (LSS) of the universe have already provided
tight constraints on cosmological parameters \citep[e.g.,][]{wang02}.
By using these constraints we can predict how much SZ power ought to
be seen at the CBI and BIMA multipole bands.
By doing so we can see if an SZ interpretation of the small-scale
fluctuations is consistent with cosmological models favored by CMB or LSS. 

In this {\em Letter} we estimate the level of the SZ angular 
power spectrum expected from cosmological models consistent with
CMB data at $l < 2000$, and compare 
it with the CBI and BIMA data at $2000<l<10000$.
We use only CMB data and a prior on the Hubble constant and do not include
any constraints from LSS.
For the primary CMB data, we use the Markov Chain Monte Carlo (MCMC) 
results of \citet{lewis02} as an estimate of the appropriate weighting 
for each cosmological model.
For the SZ effect, we use a model for the SZ power spectrum as a 
function of cosmology from \citet{komatsu02}. 
In \S\ref{sec:cmb} we outline salient features of the MCMC results.
In \S\ref{sec:sz} we sketch out our calculation of the SZ power spectrum,
including the effects of the non-Gaussian nature of the SZ fluctuations.
In \S\ref{sec:results} we compare MCMC realizations of the SZ power spectrum 
weighted by the primary CMB data with the CBI and BIMA data,
and check for consistency between them.
In \S\ref{sec:disc} we discuss implications for the 
reported detections of power at small angular scales.

\section{CMB-Calibrated Cosmological Models}\label{sec:cmb}

As a subset of cosmological models consistent with the current CMB data, 
we use the MCMC results of \citet{lewis02}
(for details see http://cosmologist.info/cosmomc). 
Estimating cosmological parameters using MCMC methods \citep{christensen01} 
entails taking random steps 
in parameter space, accepting a step if the new point is more likely,
otherwise accepting the step with some probability (less than one) set 
by how much worse the new point is than the current point. 
The resulting list of points should sample the likelihood function in the
multidimensional parameter space.

An advantage of MCMC methods (besides often being faster than grids)
is that the resulting list of consistent models (the ``chain'') provides
a self-consistent sampling of the likelihood distribution. 
A distribution of CMB-derived cosmological parameters using the MCMC chain 
therefore provides a maximally informative CMB prior. 
Covariances between parameters are naturally incorporated.
The chain provides more samples where the density is higher; 
this feature is actually a disadvantage when dealing with models lying in 
the tail of the distribution, as we will return to later. 

The chain that we use \citep{lewis02} assumes a flat universe,
no tensor component, and an equation of state for the dark energy
component of $w=-1$ (i.e., assumes a cosmological constant rather than 
quintessence).  A prior on the Hubble constant ($h=0.72 \pm 0.08$) 
consistent with the Hubble Key Project results \citep{freedman01} is
assumed, but no LSS priors.
There are six free parameters in the chain:
$\Omega_{\rm b} h^2$, $\Omega_{\rm CDM} h^2$, 
$\Omega_\Lambda$, $z_{\rm re}$, $n_{\rm s}$,
and $A_{\rm s}$, where $\Omega_{\rm b}$ is the baryon density relative to the 
critical density, $\Omega_{\rm CDM}$ the cold dark-matter density, 
$\Omega_{\Lambda}$  the cosmological-constant energy density, 
$z_{\rm re}$  the redshift of reionization, $n_{\rm s}$  the scalar 
spectral index, and $A_{\rm s}$ the amplitude of density fluctuations. 
The constraint that spatial curvature is zero defines the Hubble
constant as 
$h=\left[
   (\Omega_{\rm b}+\Omega_{\rm CDM})h^2/(1-\Omega_\Lambda)
   \right]^{1/2}$. 
The total matter density 
$\Omega_{\rm CDM}+\Omega_{\rm b}$ and $\sigma_8$ are derived 
parameters for each model.
There are 2,596 MCMC samples in total.

For each of the cosmological models in the chain, we predict
the SZ power spectrum. 
The resulting distribution of powers represents the prior probability 
of a given amount of SZ power, given the current large-angle CMB data,
i.e., CMB-normalized predictions for the SZ power spectrum.

\section{Angular Power Spectrum of the SZ Effect}\label{sec:sz}

The SZ effect arises from Compton scattering of CMB photons with
hot electrons in gas in halos \citep{sunyaev80}.
Many authors have estimated the SZ angular power spectrum
$C_l$ using analytic methods or hydrodynamic simulations, 
with broad agreement between the calculations to within a 
factor of two at $l<10^4$ 
\citep[e.g., see][for a recent analytic prediction and comparison 
between analytic methods and numerical simulations;
see  references therein for previous work]{komatsu02}.
\citet{komatsu02} find that $C_l$ is sensitive to the baryon 
density and $\sigma_8$ with an approximate scaling of 
$C_l\propto \sigma_8^7 (\Omega_{\rm b} h)^2$, but is almost 
independent of any other cosmological parameters.

We compute $C_l$ as a sum of one-halo Poisson contributions over 
all halos which can contribute:
\begin{equation}
 \label{eq:cl}
 C_l= g_\nu^2 \int dz \frac{dV}{dz}
              \int dM \frac{dn(M,z)}{dM}
	      \left|\tilde{y}_l(M,z)\right|^2,
\end{equation}
where $V(z)$ is the comoving volume of the universe at $z$ per steradian,
$dn(M,z)/dM$ the comoving dark-matter halo mass function, and
$g_\nu$ the spectral function of the SZ effect \citep{sunyaev80}.
We ignore the correlated contribution as it is unimportant at $l>300$ 
\citep{komatsu99}.
The physics of gas in halos is encoded in $\tilde{y}_l(M,z)$, 
the 2D Fourier transform of the Compton $y$-parameter
as a function of halo mass and redshift.  
The $y$-parameter is directly proportional to the integrated gas 
pressure along the line of sight; this leads to 
$C_l \propto (\Omega_{\rm b}h)^2$.

We use the method of \citet{komatsu02} for computing $C_l$. 
In brief, we compute a gas-pressure profile using the universal
gas-density and temperature profiles derived by \citet{komatsu01}
which make three assumptions about gas in halos: 
(1) hydrostatic equilibrium between gas pressure and the gravitational 
force induced by a universal dark-matter density profile \citep{navarro97},
(2) the gas density tracing the dark matter density at large radii, as
observed in simulated galaxy clusters \citep[e.g.,][]{frenk99}, and
(3) a constant polytropic equation of state for the gas.
For the mass function we use the mass function of \citet{jenkins01}.
This prescription has {\em no} free parameters, and is in broad 
agreement with SZ simulations to within a factor of two at $l<10^4$, 
agreeing with simulations at least as well as different simulations 
agree with each other.

Gas cooling, energy feedback, and star formation are not included in our
models; \citet{komatsu02} have argued that these effects are not very 
important for $C_l$, which is dominated by gas outside the core of halos.
Recently, \citet{white02} have used hydrodynamic simulations to show that 
these effects affect $C_l$ by no more than a factor of two.
These studies suggest that our theoretical prediction for $C_l$ is 
accurate to better than a factor of two, and thus MCMC-derived predictions 
for $C_l$ in the next section can be trusted up to this accuracy.

The SZ fluctuations are highly non-Gaussian, as characterized by large
skewness \citep{seljak01,zhang02,white02}; thus, sampling variance 
of $C_l$ for the SZ fluctuations differs from that for Gaussian
fluctuations, and measurements of $C_l$ in adjacent bins are highly
correlated \citep{cooray01}.
We have to include the non-Gaussianity in the error analysis to 
interpret observational SZ data correctly. 
We do this by explicitly including those trispectrum configurations
which contribute to the power-spectrum covariance matrix.
The error of $C_l$ in a bin of size $\Delta l$ is given by
\begin{equation}
 \label{eq:err}
  \left(\Delta C_l\right)^2=
  f_{\rm sky}^{-1}
  \left[ \frac{2\left(C_l+C_l^{\rm N}\right)^2}{(2l+1)\Delta l}
   + \frac{T_{ll}}{4\pi} \right],
\end{equation}
where $T_{ll}$ is the angular trispectrum relevant for the non-Gaussian
sampling variance \citep{cooray01},
\begin{equation}
 \label{eq:tl}
 T_{ll}= g_\nu^4 \int dz \frac{dV}{dz}
              \int dM \frac{dn(M,z)}{dM}
	      \left|\tilde{y}_l(M,z)\right|^4.
\end{equation}
The expression for $\Delta C_l$ is correct when both $C_l$ and 
$T_{ll}$ are sufficiently smooth in $l$ as is the case here.
\citet{komatsu02} have found that the predicted $\Delta C_l$ agrees with
hydrodynamic simulations to within a factor of two.

For the instrumental-noise power spectrum $C_l^{\rm N}$
we use $l(l+1)C_l^{\rm N}/(2\pi)=500$, 1000, 1500, and $3000~\mu{\rm K}^2$ 
at $l=1703$, 2183, 2630, and 3266 for CBI, and 720 and 2000~$\mu{\rm K}^2$ 
at $l=5237$ and 8748 for BIMA.
For sky coverage of observations $4\pi f_{\rm sky}$ we use
$1~{\rm deg}^2$ for CBI, and $0.1~{\rm deg}^2$ for BIMA.
The bin sizes are $\Delta l=565$, 378, 612, 1000, 2870, and 4150.
Although we do not use the exact window functions for the experiments, 
$C_l$ is very flat at these multipoles, reducing the importance of the 
window functions. 

\section{Results}\label{sec:results}

In figure 1 we show the histogram (weighted according to the weights
of \citet{lewis02} for each model) of expected SZ power for CBI and 
BIMA high-$l$ bins, where the three highest-$l$ CBI bins and the
two BIMA bins are averaged to single bins for each experiment.
The most likely values for the expected SZ power are less than 
$100~\mu{\rm K}^2$.
The weighted mean values are 97 and $86~\mu{\rm K}^2$ for CBI and BIMA, 
respectively, while the median values are 79 and $69~\mu{\rm K}^2$.
We also find that the analytic approximate scaling of \citet{komatsu02}
at $l\sim 2000-6000$, $l(l+1)C_l/(2\pi) \simeq 
330~\mu{\rm K}^2~\sigma_8^7 ({\Omega_{\rm b} h / 0.035})^2$,
gives similar results; the weighted mean is $100~\mu{\rm K}^2$,
while the median is $88~\mu{\rm K}^2$.
\citet{bond02} find $20-30\%$ lower normalization for the
scaling relation at the CBI band, slightly shifting the histogram
leftward.


\centerline{{\vbox{\epsfxsize=8cm\epsfbox{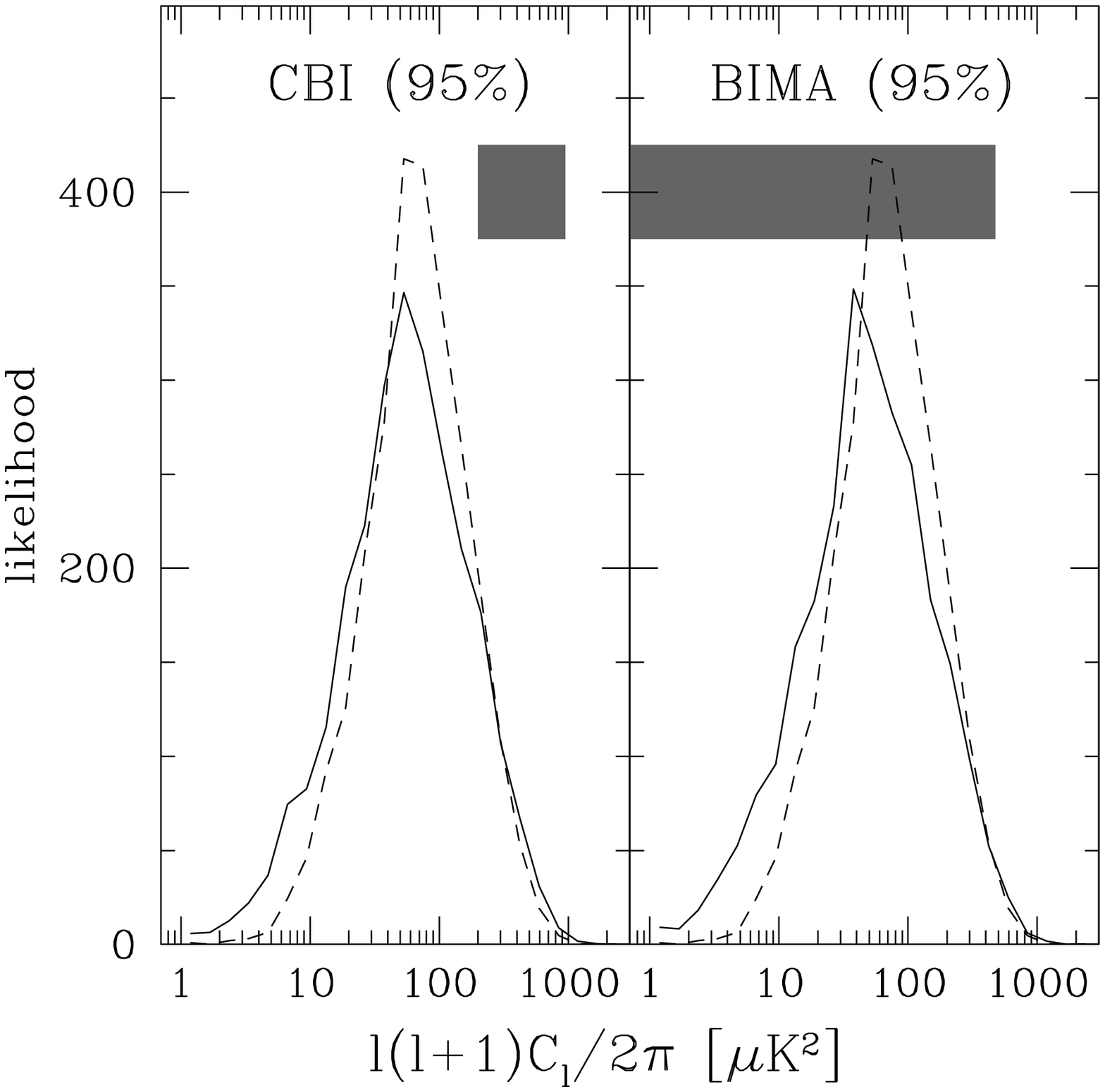}}}}
\figcaption{%
 Unnormalized probability distribution of the SZ angular power spectrum in
 CBI (left panel) and BIMA (right panel) high-$l$ averaged windows
 (solid lines), weighted by the primary CMB likelihood.
 The gray areas show 95\% confidence regions of the CBI \citep{mason02}
 and BIMA \citep{dawson02} data, assuming Gaussian fluctuations.
 Note that the true error is larger because of non-Gaussian nature of the 
 SZ fluctuations, and depends on cosmological models.
 Also shown is the result using analytic approximation of 
 \citet{komatsu02} for $C_l$ (dashed lines).
  }%
\vspace{0.5cm}

For CBI the 95\% lower limit (assuming Gaussian fluctuations) to 
the observed anisotropy is $199~\mu{\rm K}^2$, 
making an SZ interpretation of the CBI measurement only marginally 
consistent with the CMB-calibrated theories. 
We show this explicitly in figure 2 where we plot the number of 
standard deviations of the CBI and BIMA data relative to the predicted
levels of SZ fluctuations.
We have computed the errors $\Delta C_l$ using equation~(\ref{eq:err}).
We find that the CBI measurements at $l=2183$ and 2630 are inconsistent 
with most of the models that are consistent with the primary CMB data 
at greater than 2.5 and $2\sigma$ levels, respectively.
The rest of the CBI bins and the BIMA bins are consistent 
with the predicted levels of the SZ effect.

The cutoff at the $2.5-3\sigma$ level in figure 2 is due to instrumental
noise.  At the $2.5-3\sigma$ level the current data are consistent with
no SZ signal.  Thus, lower noise experiments could extend the 
histograms to larger $\sigma$ values.
Note that the largest values of $\sigma$ in the figure are higher 
than signal-to-noise ratios of CBI quoted by \citet{mason02}.
This is because $\Delta C_l$ depends on the SZ power spectrum;
a lower SZ power spectrum leads to lower
sample variance, and therefore a  smaller $\Delta C_l$.
As a result, at the largest $\sigma$ the instrumental noise determines
$\Delta C_l$ entirely, making it smaller than the quoted CBI errors.

Which models can account for the data at $l=2000-3000$?
We find that those models typically have $\sigma_8\ga 1$ and 
$\Omega_{\rm b} h^2$ higher than the BBN-allowed value of 
$\Omega_{\rm b} h^2 = 0.020 \pm 0.002$ \citep{burles01}.
A simple scaling of $C_l\propto \sigma_8^7\left(\Omega_{\rm b}h\right)^2$ 
accounts for this; thus, we need $\sigma_8\ga 1$ for a given 
BBN value of $\Omega_{\rm b}h$ to account for the CBI data 
by the SZ effect.
This conclusion agrees with \citet{bond02} and \citet{komatsu02}.

In principle, it would be straightforward to do a joint likelihood
analysis of the primary CMB and the SZ effect using CMB data on 
all angular scales.
In practice, the strong non-Gaussianity of the SZ signal, combined with 
the uncertain field selection effects (e.g., no bright point sources) 
make such a procedure still premature. 
Furthermore, the required $\sigma_8\ga 1$ is in the tail of the 
CMB-preferred distribution, so the resolution of the MCMC chain is
poor in this region. 
Importance sampling becomes unreliable in regions where the density 
of points is low.


\centerline{{\vbox{\epsfxsize=8cm\epsfbox{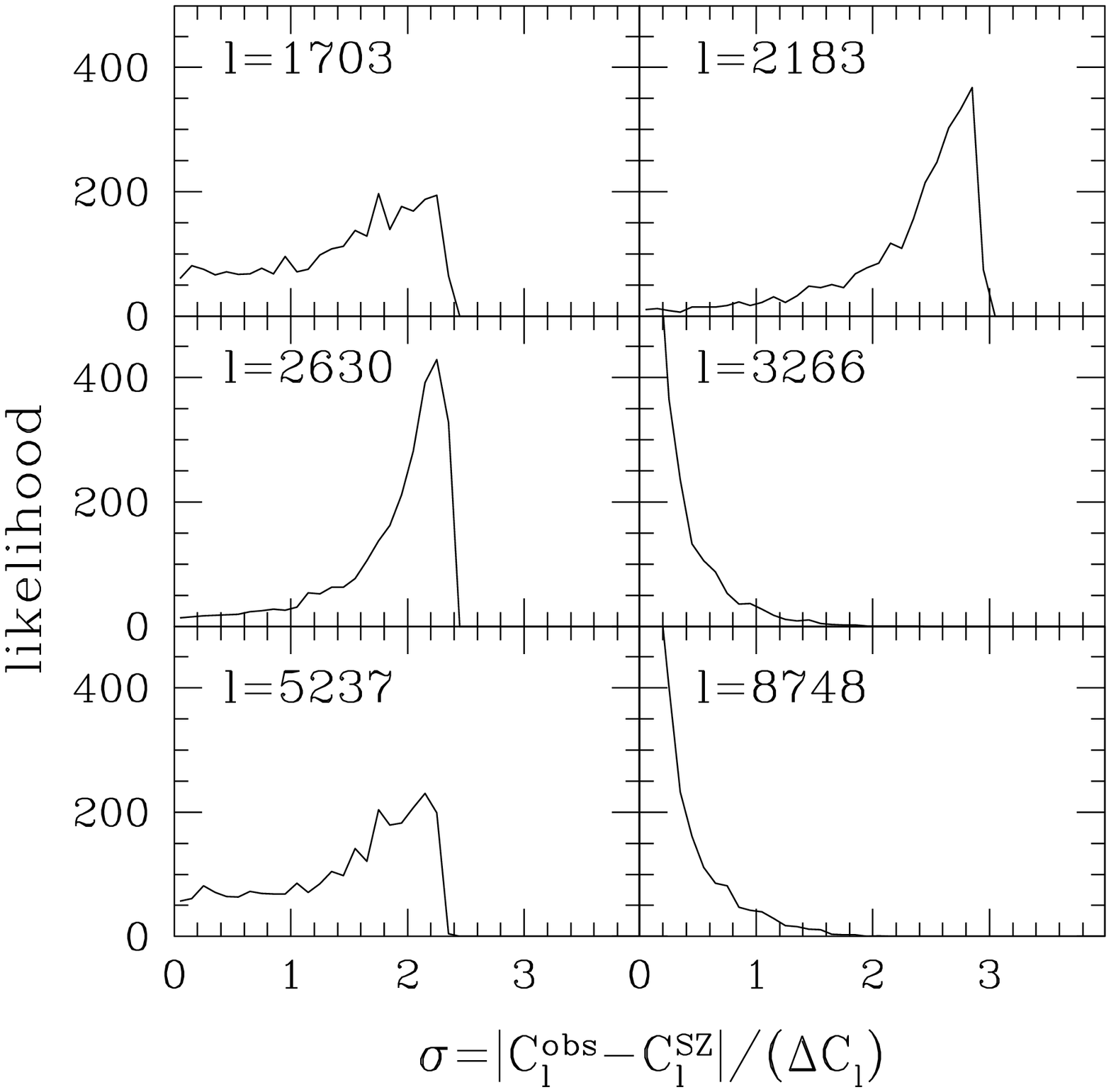}}}}
\figcaption{%
Number of standard deviations of observed data points relative to 
predicted SZ fluctuations (each data point corresponds to a panel), 
weighted by the primary CMB likelihood.
The top four panels are the CBI data points \citep{mason02}, while the
bottom two panels the BIMA data \citep{dawson02}.  
The effects of non-Gaussianity have been included in the error estimation.
}%
\vspace{0.5cm}

\section{Discussion}
\label{sec:disc}

We have presented estimates of the SZ angular power spectrum 
at $l \ga 2000$ for cosmological models that are consistent with the 
current CMB anisotropy data at $l \la 2000$. 
The mode of distribution of the predicted SZ power at $l \sim 3000$ 
is 50~$\mu{\rm K}^2$, while we find that the weighted mean is 
97~$\mu{\rm K}^2$ (the median is 78.5~$\mu{\rm K}^2$), still 
significantly lower than the reported CBI detection of anisotropy on 
these angular scales, 508~$\mu{\rm K}^2$.  
Moving out from the mode at levels of equal likelihood, 
the ranges of power enclosed by 68\% and 95\% of the distribution are 
$20-200~\mu{\rm K}^2$ and $5-450~\mu{\rm K}^2$, respectively.
This is effectively a calculation of the primary CMB prior for the SZ power
spectrum at the CBI band.
When compared to the quoted 68\% ($359-624~\mu{\rm K}^2$)
and 95\% ($199-946~\mu{\rm K}^2$) regions of the CBI detection at face value,
we find that the CBI detection is only marginally consistent with the 
SZ effect.
The BIMA detection is fully consistent with the SZ effect.

By doing a careful statistical analysis, taking into account
the non-Gaussianity of SZ fluctuations and sampling variance, we find that 
an SZ interpretation of the CBI data is inconsistent at more than 
$2-2.5\sigma$ for those cosmological models which are most 
consistent with the primary CMB data. 
The models with the highest SZ power share several characteristics: 
all models with more than 400~$\mu{\rm K}^2$ of power but a few
exceptions have $0.95<\sigma_8$ and $0.05<\Omega_{\rm b}$.
Furthermore, such models prefer a low Hubble constant ($h<0.7$) and a 
high matter density ($0.3<\Omega_{\rm CDM}+\Omega_{\rm b}<0.8$).
A reason for a low $h$ is that the primary CMB data 
tightly constrain $\Omega_{\rm b}h^2$ while we 
need a higher $\Omega_{\rm b}h$ to make the SZ effect larger. 
We can accomplish this by increasing $\Omega_{\rm b}$ while 
reducing $h$ slightly within the HST-$h$ prior.
A model with large red tilt ($n_{\rm s}<0.95$) can not produce a large 
amount of SZ power.

The possible presence of radio point sources that partially ``fill in''
the SZ effect from clusters \citep{holder02} further enhances the
possible discrepancy.
A reliable determination of $\sigma_8$ significantly lower than 1 would 
be very difficult to reconcile with an SZ interpretation of the
measured high-$l$ power.

Uncertainty in estimates for the SZ power spectrum is not yet 
fully understood, making a more detailed interpretation of the excess 
power complicated.
A major issue is that different simulations have not yet converged, 
even for adiabatic simulations.
Although our current knowledge of missing physics such as gas cooling, 
star formation, or energy feedback is still limited, 
these effects appear to not be very important \citep{white02} except on very 
small angular scales ($<1'$ or $l>10^4$) where other effects like 
non-sphericity or merging of halos may also play a role. 
Nevertheless, the current differences among analytic models, simulations, 
and estimates of the effects of missing physics are at the level of a 
factor of two in $C_l$, while the discrepancy between these predictions and 
the CBI data is about a factor of five. 

Given the strong dependence of $C_l$ on $\sigma_8$, we argue that 
the discrepancy is due to the difference between a low $\sigma_8\la 0.8$ 
favored by the primary CMB and a high $\sigma_8\ga 1$ favored by 
the SZ effect \citep{bond02,komatsu02}.
\cite{lahav02} and \cite{melchiorri02} have found similarly 
low $\sigma_8$ from the primary CMB data with the same prior on $h$.
{\em If the excess power is really due to the SZ effect}, then
this discrepancy is suggesting that there are some missing components 
in our analysis.
Multi-band SZ observations covering several frequencies 
will be required to verify the apparent discrepancy.

What is missing in our analysis?
The chain that we have used has a strong HST prior on $h$. Lower values of $h$
reduce the discrepancy, but $h \la 0.4$ would be required to explain
the entire difference.
Additional components such as massive neutrinos would make the discrepancy 
worse by driving $\sigma_8$ to even lower values. Allowing tensor modes
will have competing effects of reducing the overall normalization of
scalar modes at large scales but also allowing a blue tilt 
(higher $n_{\rm s}$), leaving the effects
on cluster scales largely unchanged. 
The effect of allowing a general equation of state for the dark energy 
will be to slightly enhance the SZ fluctuations for a fixed value of 
$\sigma_8$ \citep{komatsu02}, but to significantly reduce the 
CMB-preferred value of $\sigma_8$. 
With the SZ fluctuation power going as $\sigma_8^7$, the latter effect 
will dominate and $w>-1$ will generally reduce the expected fluctuation power.
The addition of isocurvature fluctuations may help to reconcile 
the discrepancy, while it significantly 
expands the allowed range of many cosmological
parameters \citep{trotta01}.

A very early reionization of the universe ($z_{\rm re}>20$)
will increase the CMB-preferred value of $\sigma_8$, helping to reduce 
the discrepancy.
We find a broad peak in the distribution of models
in the chain around $\sigma_8e^{-\tau}\sim 0.8$, where $\tau$
is the Thomson-scattering optical depth of the universe; 
$\tau\ga 0.22$ or $z_{\rm re}\ga 20$ would comfortably 
allow $\sigma_8\ga 1.0$. 
The discrepancy thus disappears if the universe was reionized early.
The current CMB data cannot break the degeneracy between 
$\sigma_8$ and $\tau$, and allow this area of parameter space,
although $\tau \ga 0.3$ ($z_{\rm re} \ga 30$) appears to be ruled out.
CMB-polarization experiments on large angular scales 
(e.g., MAP or Planck) should be able to break this degeneracy, 
and detect the signature of reionization 
\citep{zaldarriaga97,eisenstein99,kaplinghat02}.

We have presented an example of the ease and power of MCMC methods in applying
CMB constraints to calculations that include non-trivial dependence 
on cosmological parameters. 
It would be worthwhile to investigate the effects of LSS priors, which
should make the discrepancy worse by making $\sigma_8$
smaller \citep{lewis02,bond02}. 
With the strong preference of the high-$l$ measurements for high values of
$\sigma_8$ and $\Omega_{\rm b} h$, it would require running a new
chain that includes the CBI and BIMA data points, as importance sampling is 
unreliable in the tails of the current distribution.

Measurements of CMB anisotropy will continue to improve. 
MCMC methods provide a natural way to incorporate strong constraints 
on cosmological parameters from CMB experiments into other cosmological 
studies.
Using CMB information to understand the effects of cosmology will allow 
better understandings of systematic errors in the measurements and 
insight into important astrophysical processes.

\acknowledgements{%
 Many of the calculations and some of the text were generously provided
 by Eiichiro Komatsu, of Princeton University, who could not be a co-author 
 due to considerations of a possible conflict of interest but was an 
 important contributor.
 We would like to thank David N. Spergel and Uros Seljak 
 for useful discussions.
 GPH is supported by the W. M. Keck Foundation at the IAS.
 We are grateful to Sarah Bridle and Antony Lewis for generously providing 
 their MCMC results publicly.
}%

\end{document}